\documentclass[aps,floats,showpacs,nofootinbib,floatfix]{revtex4}%
\usepackage{epsfig}
\usepackage{amsfonts}
\usepackage{amsmath}
\usepackage{graphicx}
\usepackage{amssymb}%
{\catcode`\|=\active \gdef\Braket#1{\left<\mathcode`\|"8000\let|\bravert
{#1}\right>}}
\def\bravert{\egroup\,\vrule\,\bgroup}

\begin{document}
\title{The pion transition form factor and the pion distribution amplitude.}
\author{S. Noguera}
\email{Santiago.Noguera@uv.es}
\author{V. Vento}
\email{Vicente.Vento@uv.es}
\affiliation{Departamento de Fisica Teorica and Instituto de F\'{\i}sica Corpuscular,
Universidad de Valencia-CSIC, E-46100 Burjassot (Valencia), Spain.}
\date{\today }

\begin{abstract}
Recent BaBaR data on the pion transition form factor, whose $Q^{2}$ dependence
is much steeper then predicted by asymptotic Quantum Chromodynamics (QCD),
have caused a renewed interest in its theoretical description. We present here
a formalism based on a model independent low energy description
and a high energy description based on QCD, which match at a scale $Q_{0}$.
The high energy description incorporates a flat pion distribution amplitude, $\phi\left(  x\right)  =1,$ 
at the matching scale $Q_{0}$ and QCD evolution from $Q_{0}$ to $Q>Q_{0}$.
The flat pion distribution is connected, through soft pion theorems and
chiral symmetry, to the pion valance parton distribution at the same low scale
$Q_{0}$.  The procedure leads to  a good description of the data, 
and incorporating additional twist three
effects, to an excellent description of the data.

\end{abstract}

\maketitle

%\pacs{12.38.Lg, 12.39.St, 13.40.Gp, 13.60.Le}

%12.38.Lg 	Other nonperturbative calculations 
%12.39.St 	Factorization 
%13.40.Gp 	Electromagnetic form factors 
%13.60.Le 	Meson production 

\section{Introduction}

The implications of recent data by the BaBar Collaboration \cite{:2009mc} on
the transition form factor $\gamma^{\ast}\gamma\rightarrow\pi^{0}$, in our
understanding of the structure of the pion are being widely discussed
\cite{Radyushkin:2009zg,Polyakov:2009je,Mikhailov:2009sa,Dorokhov:2009zx,Kochelev:2009nz}%
. These results have cast doubts on our understanding on the behavior, as a
function of the light-cone momentum fraction $x$, of the pion distribution
amplitude($\pi DA$) $\phi_{\pi}(x)$ \cite{Efremov:1979qk,Chernyak:1981zz}.
Some investigations have produced scenarios where the pion $\pi DA$ is flat,
i.e. a constant value for all $x$ \cite{Radyushkin:2009zg,Polyakov:2009je},
and are in good agreement with the data for the form factor. These scenarios
are compatible with QCD sum rules \cite{Chernyak:1981zz} and lattice QCD
\cite{DelDebbio:2005bg,Braun:2006dg} calculations which provide values for the
second moment of the $\pi DA$ which are large compared with the asymptotic
value $6x(1-x)$ \cite{Efremov:1979qk} \footnote{We normalize our $\pi DA$ to
1.}. Model calculations, Nambu-Jona-Lasinio (NJL)
\cite{Praszalowicz:2001wy,RuizArriola:2002bp,Courtoy:2007vy} and the
"spectral" quark model \cite{RuizArriola:2003bs} give a constant pion
distribution amplitude, $\phi(x)=1$. The pion transition form factor calculated in these models,
however, overshoots the data
\cite{Broniowski:2009ft}.

In here we present a formalism to calculate the pion transition form factor
($\pi TFF$) which provides an excellent description in the whole range of
experimental data. The formalism consists of three ingredients: \textit{\ i})a
low energy description of the $\pi TFF$; \textit{\ ii}) a high energy
description of the $\pi TFF$; \textit{\ iii}) a matching condition between the
two descriptions at a scale $Q_{0}$ characterizing the separation between the
two regimes. For the low energy description we take a parametrization
of the low energy data to avoid model dependence at $Q_{0}$. 
The high energy description of the $\pi TFF$, defined by the pion
Distribution Amplitude ($\pi DA$), contains Quantum Chromodynamic (QCD)
evolution from $Q_{0}$ to any higher $Q$, the recently introduced mass cut-off
procedure \cite{Radyushkin:2009zg}, and ultimately additional higher twist
effects. The relation between the leading twist ($\pi DA$) and the pion Parton
Distribution ($\pi PD$) allows us to use a scheme for the determination of
$Q_{0}$ that has proven successful to describe parton distributions of mesons
and baryons \cite{Davidson:1994uv,Traini:1997jz,Scopetta:1997wk,Scopetta:1999ud,Noguera:2005cc,Broniowski:2007si,Courtoy:2008nf}. 
The scheme consists in evolving the second moments of the parton
distributions from a scale where they are experimentally known to a scale
$Q_{0}$ where they coincide with those determined by a low energy description.
Once $Q_{0}$ is known, higher $Q$ results for any other observable are
obtained by evolving the low energy description of the corresponding
observable from $Q_{0}$ to the required scale $Q$. Our analysis will
systematically explore the region of validity of the scheme and find
implications of new physics in the intermediate energy region.

In section II we define the $\pi TFF$ both for low and high energies and
establish the matching condition at the hadronic scale $Q_{0}$ . The matching
condition of section II allows to determine the mass parameter of the
formalism once the hadronic scale is known. Section III is devoted to
establish a relation between the $\pi DA$ and the $\pi PD$. This relation
allows to characterize a unique matching scale $Q_{0}$. Section IV is
dedicated to analyze the stability of the parameters and the description of
the data. We find reasonable agreement with the experimental data  for $Q^{2}$
 below $15$ GeV$^{2}$ with parameters whose values are to be expected from
physical arguments. Two matching scales are investigated in the analysis. We
realize that none of them is able to reproduce accurately the slope of the
data in the intermediate region. In section V we incorporate in a phenomenological way additional twist three effects.
They lead to a modification of the matching condition. Their overall effect is
small at high $Q$ but they are important in reproducing the slope in the
intermediate region. In this way we are able to get an excellent agreement
with the data. In section VI we draw some conclusions. Finally, in the Appendix
we present a  model calculation which overshoots the data and therefore signals
 the importance of evolution in the description of the experiment. Morevoer, it also
serves to understand the order of magnitude of some of the  parameters
used in our approach. 

\section{The pion transition form factor.}

\label{pTFF}

The form factor $F_{\gamma^{\ast}\gamma^{\ast}\pi}$ relating two photons (real
or virtual) to the pion is a very important object of study in exclusive
processes in QCD. In particular it has been used by comparing perturbative QCD
predictions with experimental data to obtain information about the shape of
the $\pi DA$\cite{Efremov:1979qk,Lepage:1980fj,Chernyak:1983ej}.
Experimentally for small virtuality of one of the photons it has been measured
by the CELLO \cite{Behrend:1990sr}, by the CLEO \cite{Gronberg:1997fj} and
recently by the BaBaR \cite{:2009mc} collaboration. The last results are
in disagreement with previous  theoretical expectations.

In lowest order of perturbative QCD, the transition form factor for the
process $\pi^{0}\rightarrow\gamma\,\gamma^{\ast}$ is given by%

\begin{equation}
Q^{2}F(Q^{2})=\frac{\sqrt{2}f_{\pi}}{3}\int_{0}^{1}\frac{dx}{x}\phi_{\pi
}(x,Q^{2}). \label{tff}%
\end{equation}
where $Q^{2}=-q^{2},$ $q_{\mu}$ is the momentum of the virtual photon,
$\phi_{\pi}\left(  x,Q^{2}\right)  $ is $\pi DA$ at the $Q^{2}$ scale and $f_\pi =0.131$ GeV. In this
expression, the $Q^{2}$ dependence appears through the QCD evolution of the
$\pi DA.$ The $\pi DA$ can be expressed in terms of the Gegenbauer polynomials
\cite{Lepage:1979zb,Mueller:1994cn} ,
\begin{equation}
\phi_{\pi}(x,Q^{2})=x\,(1-x)\sum_{n(even)=0}^{\infty}a_{n}\,C_{n}%
^{3/2}(2x-1)\left(  \log\frac{Q^{2}}{\Lambda_{QCD}^{2}}\right)  ^{-\gamma_{n}%
}, \label{pDA0}%
\end{equation}
where $\gamma_{n}$ are the anomalous dimensions%

\begin{equation}
\gamma_{n}=\frac{C_{F}}{\beta}\left(  1+4\sum_{k=2}^{n+1}\frac{1}{k}-\frac
{2}{(n+1)(n+2)}\right)  ,
\end{equation}
$\beta=\frac{11N_{C}}{3}-\frac{2N_{f}}{3}$ is the beta function to lowest
order and $C_{F}=\frac{N_{C}^{2}-1}{2N_{C}}.$ 
Assuming that the $a_{n}$ coefficients are known, Eq. (\ref{tff}) gives the
$\pi TFF$ for any value of $Q$ high enough compared with $\Lambda_{QCD}.$ The
$a_{n}$ coefficients are obtained  if  we know  the $\pi DA$, 
$\phi_{\pi}\left(  x,Q_{0}^{2}\right)$,  at some momentum $Q_0$, using the orthogonality 
relations of the Gegenbauer polynomials%
\begin{equation}
a_{n}\left(  \log\frac{Q_{0}^{2}}{\Lambda_{QCD}^{2}}\right)  ^{-\gamma_{n}%
}=4\frac{2n+3}{\left(  n+1\right)  \left(  n+2\right)  }\int_{0}^{1}%
dx\,C_{n}^{3/2}(2x-1)\,\phi_{\pi}\left(  x,Q_{0}^{2}\right)  \label{pDA1}%
\end{equation}

In order to prevent divergences in the integrand of Eq. (\ref{tff}) it has
been assumed that $\phi_{\pi}\left(  x\right)  $ vanishes for $x=0$. There are
no fundamental reasons for this restriction, moreover, several chiral models
predict for massless pions $\phi_{\pi}(x)=1$
\cite{Praszalowicz:2001wy,RuizArriola:2002bp,Courtoy:2007vy,CourtoyThesis}. We
use this model prediction as input at $Q_{0}$ in Eq (\ref{pDA1}) and we obtain
for the expansion of the $\pi DA$ 
\begin{equation}
\phi_{\pi}(x,Q^{2})=4x(1-x)\sum_{n(even)=0}^{\infty}\frac{2n+3}{(n+1)(n+2)}%
C_{n}^{3/2}(2x-1)\left(  \frac{\log{(Q^{2}/\Lambda_{QCD}^{2})}}{\log
{(Q_{0}^{2}/\Lambda_{QCD}^{2})}}\right)  ^{-\gamma_{n}}\ .\label{pDA}%
\end{equation}
This expression has the right asymptotic behavior, $\lim_{Q\rightarrow\infty
}\phi_{\pi}(x,Q)=6x\left(  1-x\right)  $. Nevertheless, despite its
appearance, the $\pi DA$ defined in Eq. (\ref{pDA}) has the properties
$\lim_{x\rightarrow0^{+}}\phi_{\pi}(x,Q_{0})=\lim_{x\rightarrow1^{-}}\phi
_{\pi}(x,Q_{0})=1.$ In order to cure the divergences in the integrand of Eq.
(\ref{tff}) we follow the proposal of Radyushkin in section II.B of ref.
\cite{Radyushkin:2009zg}, and we introduce a cutoff mass $M$ in the above
definition of the $\pi TFF$, in terms of the $\pi DA$,%
\begin{equation}
Q^{2}F(Q^{2})=\frac{\sqrt{2}f_{\pi}}{3}\int_{0}^{1}\frac{dx}{x+\frac{M^{2}%
}{Q^{2}}}\phi_{\pi}(x,Q^{2}).\label{tff_R}%
\end{equation}
We will discuss later the determination of the  $M.$%

\begin{figure}
[ptb]
\begin{center}
\includegraphics[
height=4.0967cm,
width=10.2516cm
]%
{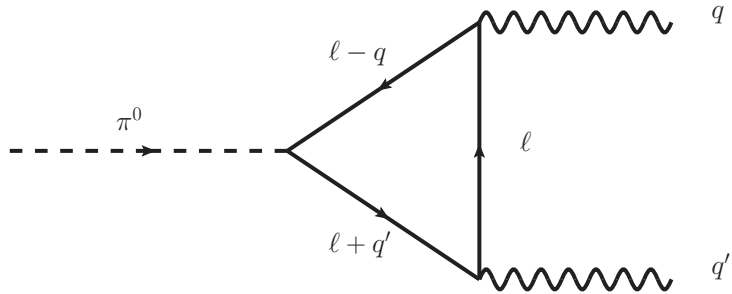}%
\caption{Calculation of the transition form factor via the triangle diagram.}%
\label{Trianglediagrams}%
\end{center}
\end{figure}

At low $Q^{2}$ we can parametrize the data as \cite{Amsler:2008zzb},
\begin{equation}
F\left(  Q^{2}\right)  =F\left(  0\right)  \left[  1-a\frac{Q^{2}}{m_{\pi^{0}%
}^{2}}+...\right]  \label{ExpFit}%
\end{equation}
with $F\left(  0\right)  =0.273(10)%
%TCIMACRO{\unit{GeV}}%
%BeginExpansion
\operatorname{GeV}%
%EndExpansion
^{-1}$ and $a=0.032\left(  4\right)  ,$ constants which have been obtained
from the experimental study of $\pi^{0}\rightarrow\gamma\,e^{+}\,e^{-}$. The
$\pi TFF$ at these low values of $Q^{2}$ can be described by several hadron
models. For instance, the parametrization can be well understood studying the
triangle diagram (see Fig. \ref{Trianglediagrams}). This calculation is
straightforward (see appendix) leading, in the case of massless pions, to
$F(0)=\sqrt{2}/\left(  4\pi^{2}f_{\pi}\right)  =0.273%
%TCIMACRO{\unit{GeV}}%
%BeginExpansion
\operatorname{GeV}%
%EndExpansion
^{-1},$ 
which coincides with the experimental value. The  parameter $a$ can be linked
to the constituent quark mass, obtaining $m_{q}=m_{\pi}/\sqrt{12\,a}\simeq0.220%
%TCIMACRO{\unit{GeV}}%
%BeginExpansion
\operatorname{GeV}%
%EndExpansion
$.

Despite the fact that several models reproduce the low energy data, in order
to have a model independent expression for the form factor, we adopt a
monopole parametrization of the $\pi TFF$ in the low energy region:
\begin{equation}
F^{LE}\left(  Q^{2}\right)  =\frac{F\left(  0\right)  }{1+a\frac{Q^{2}}%
{m_{\pi^{0}}^{2}}}\,. \label{tff_EXP}%
\end{equation}

We have thus introduced two descriptions for the $\pi TFF$, one in the low energy
regime defined by $F^{LE}(Q^{2})$, and one in the high energy regime, defined by Eqs.(\ref{pDA})
and (\ref{tff_R}). Both descriptions will be matched at $Q_{0}$, the scale
previously introduced which we now define. If $Q>Q_{0}$ the pion form factor
should be calculated using Eqs. (\ref{pDA}) and (\ref{tff_R}) , and for
$Q<Q_{0}$ by Eq. (\ref{tff_EXP}). For $Q=Q_{0}$ both expressions should lead
to the same result. In this way $Q_{0}$ is defined as the matching point
between the low and high energy descriptions. Consistency implies that this
matching point must be the same for any physical quantity.

Note that our formalism contains two unknowns, $Q_{0}$ and $M$. The scale
$Q_{0}$ is an important ingredient of our calculation
\cite{Traini:1997jz,Scopetta:1997wk,Scopetta:1999ud} and determines the range
of validity of the low energy description used in the calculation, i.e. for
$Q<Q_{0}$ $\phi_{\pi}(x)=1$ and $F^{LE}\left(  Q^{2}\right)  $ is assumed to
be a good representation of the theory. $Q_{0}$ is fixed by the value of
$\Lambda_{QCD}$, and the physical results should not show any dependence on
it. $M$ contains not only the effects associated with the constituent quark
mass but, as pointed out by Radyushkin \cite{Radyushkin:2009zg}, also the
contribution of some mean transverse momentum.

 We next equate the pion form factor calculated via equation (\ref{tff_R})
at $Q^{2}=Q_{0}^{2}$, where $\phi_{\pi}(x,Q_{0})=1$,%
\begin{equation}
Q_{0}^{2}F(Q_{0}^{2})=\frac{\sqrt{2}f_{\pi}}{3}\int_{0}^{1}\frac{dx}%
{x+\frac{M^{2}}{Q_{0}^{2}}}=\frac{\sqrt{2}f_{\pi}}{3}ln\frac{Q_{0}^{2}+M^{2}%
}{M^{2}},
\end{equation}
to the value given by the monopole parametrization Eq. (\ref{tff_EXP}),
\begin{equation}
\frac{\sqrt{2}f_{\pi}}{3}ln\frac{Q_{0}^{2}+M^{2}}{M^{2}}=Q_{0}^{2}F^{LE}%
(Q_{0}^{2})=\frac{F\left(  0\right)  \,Q_{0}^{2}}{1+a\frac{Q_{0}^{2}}%
{m_{\pi^{0}}^{2}}}. \label{cont}%
\end{equation}

This is the first equation relating the two parameters. We proceed in the next
section to find a second relation relation which allows a full determinations
of the parameters.

\section{The pion distribution amplitude and the pion parton distribution.}

In this section we establish a relation between the $\pi DA$ and the pion
valence parton distribution ($\pi VPD$) which will be instrumental in the
determination of $Q_{0}$.

The $\pi DA$ is defined by,%

\begin{equation}
\delta_{ij}\,i\,\sqrt{2}\,f_{\pi}\ \phi_{\pi}\left(  x\right)  =\int
\frac{d\lambda}{2\pi}~e^{i\lambda\,x}~\left\langle 0\right\vert \bar{\psi
}\left(  0\right)  \gamma^{+}\gamma_{5}\tau^{i}\psi\left(  \lambda n\right)
\left\vert \pi^{j}\left(  P\right)  \right\rangle . \label{DefpDA}%
\end{equation}
In order to introduce a convenient expression for the valence parton
distribution let us define the quark distribution
\begin{equation}
H_{q}^{i,j}\left(  x\right)  =\frac{1}{2}\int\frac{d\lambda}{2\pi}%
e^{i\lambda\,x}\,\left\langle \pi^{i}\left(  P\right)  \right\vert \bar{\psi
}\left(  0\right)  \,\gamma^{+}\,\frac{1}{2}\left(  1+\alpha\,\tau^{3}\right)
\,\psi\left(  \lambda n\right)  \left\vert \pi^{j}\left(  P\right)
\right\rangle =\frac{1}{2}\left(  \delta_{i,j}\,H^{I=0}\left(  x\right)
+i\epsilon_{i3j}\,\alpha\,H^{I=1}\left(  x\right)  \right)  \label{DefH}%
\end{equation}
with $\alpha=\pm1$ for $q=u,d$ and $H^{I=0,1}\left(  x\right)  $ represent the
isospin decompositions. The quark distribution is only non-vanishing in the
interval $-1\leq x\leq1.$ This quark distribution is related with the standard
$q\left(  x\right)  $ and $\bar{q}\left(  x\right)  $ distributions, by%
\begin{equation}
H_{q}^{\pi}=q\left(  x\right)  \,\theta\left(  x\right)  -\bar{q}\left(
-x\right)  \,\theta\left(  -x\right)  \ ,
\end{equation}
with
\begin{align}
H_{u}^{\pi^{+}}  &  =H_{d}^{\pi^{-}}=\frac{1}{2}\left(  H^{I=0}\left(
x\right)  +H^{I=1}\left(  x\right)  \right)  \ .\\
H_{d}^{\pi^{+}}  &  =H_{u}^{\pi^{-}}=\frac{1}{2}\left(  H^{I=0}\left(
x\right)  -H^{I=1}\left(  x\right)  \right)  \ .
\end{align}
The pion valence parton distribution ($\pi VPD$) is defined as $H_{Vq}^{\pi
}\left(  x\right)  =H_{q}^{\pi}\left(  x\right)  +H_{q}^{\pi}\left(
-x\right)  .$ Using charge conjugation, one can prove that the isoscalar and
isovector part the quark distribution, $H^{I=0,1}\left(  x\right)  ,$ are
respectively odd and even functions of $x$. Therefore%
\begin{align}
H_{Vq}^{i,j}\left(  x\right)   &  =\frac{1}{2}\int\frac{d\lambda}{2\pi
}e^{i\lambda\,x}\,\left\langle \pi^{i}\left(  P\right)  \right\vert \bar{\psi
}\left(  0\right)  \,\gamma^{+}\,\frac{1}{2}\,\alpha\,\tau^{3}\,\psi\left(
\lambda n\right)  \left\vert \pi^{j}\left(  P\right)  \right\rangle
\nonumber\\
&  +\frac{1}{2}\int\frac{d\lambda}{2\pi}e^{-i\lambda\,x}\,\left\langle \pi
^{i}\left(  P\right)  \right\vert \bar{\psi}\left(  0\right)  \,\gamma
^{+}\,\frac{1}{2}\,\alpha\,\tau^{3}\,\psi\left(  \lambda n\right)  \left\vert
\pi^{j}\left(  P\right)  \right\rangle \label{DefpHV}%
\end{align}
In terms of the standard $q$ and $\bar{q}$ distributions we have%
\begin{equation}
H_{Vq}^{i,j}\left(  x\right)  =\frac{1}{2}\alpha\,i\,\epsilon_{i3j}\,\left[
q_{V}\left(  x\right)  \theta\left(  x\right)  +q_{V}\left(  -x\right)
\theta\left(  -x\right)  \right]
\end{equation}
with$\ q_{V}\left(  x\right)  =q\left(  x\right)  -\bar{q}\left(  x\right)  .$

Soft pion theorems have been used to relate the double pion distribution
amplitude to the $\pi DA$\cite{Polyakov:1998ze}. Using the same arguments, it
is easy to connect the matrix elements of the preceeding definitions,%
\begin{equation}
\lim_{P\rightarrow0}\left\langle \pi^{i}\left(  P\right)  \right\vert
\bar{\psi}_{f}\left(  0\right)  \,\gamma^{+}\,\tau^{3}\,\psi_{f}\left(
\lambda n\right)  \left\vert \pi^{j}\left(  P\right)  \right\rangle
=i\,\epsilon_{i3\ell}\,\frac{\sqrt{2}}{i\,f_{\pi}}\,\lim_{P\rightarrow
0}\,\left\langle 0\right\vert \bar{\psi}_{f}\left(  0\right)  \gamma^{+}%
\gamma_{5}\tau^{\ell}\psi_{f}\left(  \lambda n\right)  \left\vert \pi
^{j}\left(  P\right)  \right\rangle \,. \label{SPT}%
\end{equation}
and therefore%
\begin{equation}
H_{Vq}^{\pi}\left(  x\right)  =\frac{1}{2}\left(  \alpha\,\right)
\,i\,\epsilon_{i3j}\,\left[  \phi_{\pi}\left(  x\right)  \theta\left(
x\right)  +\phi_{\pi}\left(  -x\right)  \theta\left(  -x\right)  \right]  .
\end{equation}
Substituiting Eq. (\ref{SPT}) in Eq. (\ref{DefpHV}) we obtain
\begin{equation}
\phi_{\pi}(x)=q_{V}(x), \label{phi=q}%
\end{equation}
a relation which is important for our scheme because it provides a universal
determination of the hadronic scale $Q_{0}$.

The previous result, Eq.(\ref{phi=q}), is confirmed by chiral models for the
pion, like the NJL model or the spectral quark model. In fact, these local
chiral quark models give, in the chiral limit, $\phi_{\pi}(x)=q(x)=1$. Using
the results obtained in the NJL model for the $\pi DA$ (see
\cite{Courtoy:2007vy}) and the $\pi PD$ (see \cite{Theussl:2002xp}) we obtain%
\begin{equation}
\phi_{\pi}\left(  x\right)  =1+\mathcal{O}\left(  m_{\pi}^{2}\right)
\,,\ \ \ \ \ \ q_{V}\left(  x\right)  =\frac{g_{\pi qq}\,f_{\pi}}{\sqrt
{2}\,m_{q}}+\mathcal{O}\left(  m_{\pi}^{2}\right)  =1+\mathcal{O}\left(
m_{\pi}^{2}\right)  \label{Ch_DA_PA}%
\end{equation}
where in the last relation $m_{q}$ is the constituent quark mass, $g_{\pi qq}$
is the pion quark coupling constant and the last step follows from the
Goldberger-Treiman relation. This result for $q_{V}(x)$ arises naturally from
the physical meaning of the $\pi VPD$. The flatness of the distribution
implies that the probability of finding a quark of any relative momentum
fraction $x$ is the same. This is so because in the chiral limit quarks and
pions are massless and therefore there is no mass scale to which the momentum
distribution can attach. The appearance of flat distributions is a fundamental
consequence of soft pion theorems and the chiral limit, not a peculiarity of
some models.

Since the origin of the flat distribution is related to soft pion theorems and
the chiral limit, this behavior will change as we move towards higher energy.
Starting from $Q_{0}$, the momentum scale in which the flat distributions are
a good approximation, $\phi_{\pi}(x,Q_{0})=q_{V}(x,Q_{0})=1,$ we apply the
evolution equations in order to obtain these distributions at higher $Q$
values. The $\pi VPD,$ $q_{V}(x,Q)$ evolves through the DGLAP equation, and it
concentrates close to the $x\sim0$ region for very high values of $Q$. The
$\pi DA$ evolves through the ERBL equation and, therefore, for high $Q$ values
we obtain the asymptotic result $\phi_{\pi}(x,Q)\sim6x\left(  1-x\right)  $.
In conclusion, QCD evolution obscures the relation between these two
distributions, which become distinct from one another. It follows from the
above discussion that it is unreasonable to assume that at relatively low
energies $\phi_{\pi}(x)$ is close to its asymptotic value, whereas the $\pi
VPD$ is far from it.

The evolution of the flat $\pi VPD,$ $q(x,Q_{0})=1,$ has been previously
studied in refs. \cite{Davidson:1994uv, Noguera:2005cc, Broniowski:2007si,Courtoy:2008nf}
obtaining a very good description of the experimental data. We sketch here the
procedure to fix the value of the matching point, $Q_{0}.$ We know that the
momentum fraction of each valence quark at $Q=2%
%TCIMACRO{\unit{GeV}}%
%BeginExpansion
\operatorname{GeV}%
%EndExpansion
$ is 0.23 \cite{Sutton:1991ay}. We fit the initial point of the evolution,
$Q_{0},$ imposing that the evolved $\pi VPD$ reproduces this momentum fraction
at $Q=2%
%TCIMACRO{\unit{GeV}}%
%BeginExpansion
\operatorname{GeV}%
%EndExpansion
.$ Obviously, the resulting $Q_{0}$ value will depend on the value of
$\Lambda_{QCD}$. Varying $\Lambda_{QCD}$ form $0.174%
%TCIMACRO{\unit{GeV}}%
%BeginExpansion
\operatorname{GeV}%
%EndExpansion
$ to $0.326%
%TCIMACRO{\unit{GeV}}%
%BeginExpansion
\operatorname{GeV}%
%EndExpansion
$ we obtain that $Q_{0}$ changes from $0.290%
%TCIMACRO{\unit{GeV}}%
%BeginExpansion
\operatorname{GeV}%
%EndExpansion
$ to $0.470%
%TCIMACRO{\unit{GeV}}%
%BeginExpansion
\operatorname{GeV}%
%EndExpansion
$, in leading order (LO) evolution. A change in $\Lambda_{QCD}$ implies a
change in the initial point $Q_{0}$ but, once we have fitted $Q_{0}$ through
the momentum fraction carried by the quarks at $Q=2%
%TCIMACRO{\unit{GeV}}%
%BeginExpansion
\operatorname{GeV}%
%EndExpansion
,$ the evolved parton distribution is independent of the value of
$\Lambda_{QCD}$ and the quality in the description of the experimental data is preserved.

At this point we must return to Eq. (\ref{cont}) regarding the stability of
$M.$ We obtain that $M$ changes only a few $%
%TCIMACRO{\unit{MeV}}%
%BeginExpansion
\operatorname{MeV}%
%EndExpansion
$ when we vary $\Lambda_{QCD}$ form $0.174%
%TCIMACRO{\unit{GeV}}%
%BeginExpansion
\operatorname{GeV}%
%EndExpansion
$ to $0.326%
%TCIMACRO{\unit{GeV}}%
%BeginExpansion
\operatorname{GeV}%
%EndExpansion
.$ The largest incertitude comes from the experimental error in $a,$ obtaining
$M=0.466\pm0.006%
%TCIMACRO{\unit{GeV}}%
%BeginExpansion
\operatorname{GeV}%
%EndExpansion
.$ We thus obtain a very stable value for $M$, showing the consistency of the
procedure. For the numerical analysis we have chosen $\Lambda_{QCD}=0.226%
%TCIMACRO{\unit{GeV}}%
%BeginExpansion
\operatorname{GeV}%
%EndExpansion
,$ $Q_{0}=0.355%
%TCIMACRO{\unit{GeV}}%
%BeginExpansion
\operatorname{GeV}%
%EndExpansion
$, $a=0.32$ and $M=0.466%
%TCIMACRO{\unit{GeV}}%
%BeginExpansion
\operatorname{GeV}%
%EndExpansion
.$

\section{Discussion.}

 Summarizing  the ideas developed previously, a model independent scheme of evaluation for
the $\pi TFF$ at high $Q$ values arises under the following
assumptions:\textit{\ i}) we consider, on the basis of chiral symmetry and
soft pion theorems, that at some point $Q_{0}$ the $\pi DA$ and the $\pi VPD$
coincide and have flat behavior $\phi_{\pi}(x,Q_{0})=q_{V}(x,Q_{0})=1;$
\textit{ii}) we fix $Q_{0}$ applying QCD evolution to the $\pi VPD$ from
$Q_{0}$ to $Q=2%
%TCIMACRO{\unit{GeV}}%
%BeginExpansion
\operatorname{GeV}%
%EndExpansion
$ and impose that each valence quark carries a fraction of the total momentum
of the pion of 0.23; \textit{iii}) for $Q<Q_{0}$ we assume the experimental
parametrization of $F(Q^{2})$ given in (\ref{tff_EXP}); and \textit{iv})
 for $Q>Q_{0}$ the $\pi TFF$ is given by the Eq. (\ref{tff_R}), where we fix
the $M$ parameter through the continuity condition Eq. (\ref{cont}).

\begin{figure}[ptb]
\begin{center}
\includegraphics[
height=8.8458cm,
width=10.5987cm
]{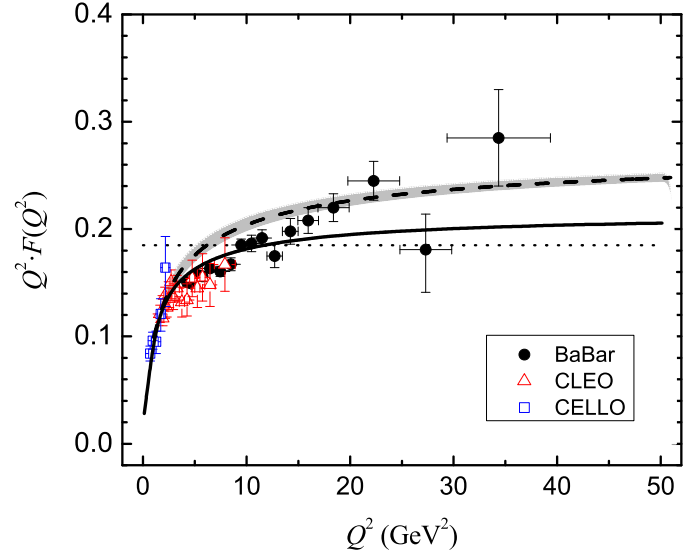}
\end{center}
\caption{Calculation of the transition form factor via the $\pi DA$ with
$M=0.466\operatorname{GeV}$, $a=0.032$ and defining the matching point at
$Q_{0}=0.355\operatorname{GeV}$ (full-line) compared with the experimental
data. The dashed-line corresponds to the same calculation defining the
matching point at $Q_{0}=1.\operatorname{GeV}$ and with
$M=0.500\operatorname{GeV}$. The dotted-line corresponds to the asymptotic
limit. The gray region gives the indeterminacy of the results due to the error
in $a,$ which is only apreciable for the $Q_{0}=1\operatorname{GeV}$ case.}%
\label{Fig2}%
\end{figure}

\begin{figure}[ptb]
\begin{center}
\includegraphics[
trim=0.000000cm 0.026365cm 0.000000cm -0.026365cm,
height=8.8458cm,
width=10.5987cm
]{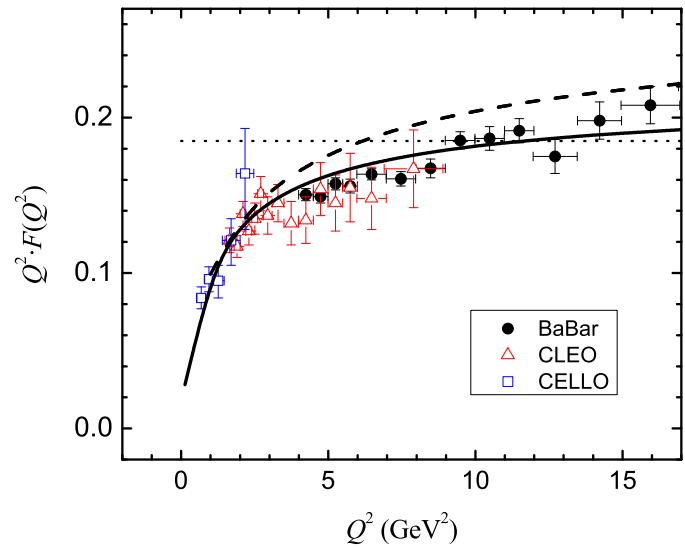}
\end{center}
\caption{Same as in Fig \ref{Fig2} for the $Q^{2}$ region under
$17\operatorname{GeV}^{2}.$}%
\label{Fig3}%
\end{figure}

In Figs. \ref{Fig2} and \ref{Fig3} we show the $\pi TFF$ calculated using the
evolved $\pi DA$ as explained above and compare with the data of refs. \cite{Behrend:1990sr,
Gronberg:1997fj, :2009mc}. 

In ref. \cite{Noguera:2005cc} for the pion, and in refs.
\cite{Scopetta:1997wk,Scopetta:1999ud} for the nucleon, it is shown that
including gluons and sea quarks at low energy the preferred starting point for
evolution is $Q_{0}\simeq1%
%TCIMACRO{\unit{GeV}}%
%BeginExpansion
\operatorname{GeV}%
%EndExpansion
$. To move $Q_{0}$ up to $1%
%TCIMACRO{\unit{GeV}}%
%BeginExpansion
\operatorname{GeV}%
%EndExpansion
$ implies to change the value of $M$ to $0.500%
%TCIMACRO{\unit{GeV}}%
%BeginExpansion
\operatorname{GeV}%
%EndExpansion
$. Dashed curves in Fig. \ref{Fig2} and \ref{Fig3} reproduces the $\pi
TFF\ $calculated with these parameters. The experimental incertitude in $a$
produces in this case a larger indetermination in the mass, obtaining
$M=0.500\pm0.040%
%TCIMACRO{\unit{GeV}}%
%BeginExpansion
\operatorname{GeV}%
%EndExpansion
.$ Nevertheless, this error does not affect to the predictive power, as shown
in Fig. \ref{Fig2}.

The calculation shows several features which we now discuss. The first effect of this model independent calculation, in comparison with
the triangle result, is to cure the asymptotic behavior of the form factor.
Model calculations overshoot the experimental data for $Q^{2} \sim
10-30\,$GeV$^{2}$ as is clear from ref. \cite{Broniowski:2009ft} and it is
confirmed by our calculation of the triangle diagram (see appendix), where we
obtain $Q^{2}F(Q^{2})\sim0.54%
%TCIMACRO{\unit{GeV}}%
%BeginExpansion
\operatorname{GeV}%
%EndExpansion
$ for $Q^{2}\sim30%
%TCIMACRO{\unit{GeV}}%
%BeginExpansion
\operatorname{GeV}%
%EndExpansion
^{2}$. Moreover, the form factor in this model calculation shows a behavior
proportional to $\left[  \log\left(  Q^{2}/m^{2}\right)  \right]  ^{2}$ for
high $Q^{2}.$ Now, with the procedure defined here, we recover the right
asymptotic result $Q^2 F\left(  Q^{2}\right)  \rightarrow\sqrt{2}f_{\pi}.$ What
is under discussion at present is not this value, but for which value of
$Q^{2}$ the asymptotic behavior is attained and the behaviour of the $\pi TFF$
in the intermediate energy region.

In Fig. \ref{Fig3} we observe that the calculation with the mass cut-off
introduced in ref. \cite{Radyushkin:2009zg} adjusts well to the data up to
$Q^{2}\sim15%
%TCIMACRO{\unit{GeV}}%
%BeginExpansion
\operatorname{GeV}%
%EndExpansion
^{2},$ when we use as matching point at $Q_{0}=0.355$ GeV. However, at that
momentum most of the new data bend upward while our calculation remains
practically flat (see Fig \ \ref{Fig2}). The slope of the $\pi TFF$ in the
region around $20%
%TCIMACRO{\unit{GeV}}%
%BeginExpansion
\operatorname{GeV}%
%EndExpansion
^{2}$ is not well reproduced. If we assume the higher matching point $Q_{0}=1%
%TCIMACRO{\unit{GeV}}%
%BeginExpansion
\operatorname{GeV}%
%EndExpansion
$, the calculation still fits relatively well the low energy data up to $5%
%TCIMACRO{\unit{GeV}}%
%BeginExpansion
\operatorname{GeV}%
%EndExpansion
^{2}$, but, contrary to the other, fails in the intermediate region, as we see
in Fig. \ref{Fig3}. In the region around $20%
%TCIMACRO{\unit{GeV}}%
%BeginExpansion
\operatorname{GeV}%
%EndExpansion
^{2},$ this calculation goes through most of the experimental points although
the slope of the curve is smaller than the one indicated by the data. The
indeterminacy in the predictions related to the experimental error in $a$ is
only apreciable when we use $Q_{0}=1%
%TCIMACRO{\unit{GeV}}%
%BeginExpansion
\operatorname{GeV}%
%EndExpansion
$ as matching point. It is represented by the gray region in Fig \ref{Fig2},
corresponding the upper limit of the region to $a=0.028$ and the lower limit
of to $a=0.036$. In both calculations, the $\pi TFF$ tends to the asymptotic
limit at very high energy, but in the studied region it is growing very slowly
($Q_{0}=1%
%TCIMACRO{\unit{GeV}}%
%BeginExpansion
\operatorname{GeV}%
%EndExpansion
$ case) or it is practically constant ($Q_{0}=0.355$ GeV case).  The value of the mass parameter, around $0.450-0.500%
%TCIMACRO{\unit{GeV}}%
%BeginExpansion
\operatorname{GeV}%
%EndExpansion
,$ is reasonable, since it receives contributions not only from de constituent
quark mass, but also from the mean value of the transverse momentum. 

\section{Possible twist three effects in the pion transition form factor.}

Using $Q_{0}=0.355%
%TCIMACRO{\unit{GeV}}%
%BeginExpansion
\operatorname{GeV}%
%EndExpansion
$ as matching point, the experimental results are well reproduced up to $15%
%TCIMACRO{\unit{GeV}}%
%BeginExpansion
\operatorname{GeV}%
%EndExpansion
^{2},$ but it is difficult to accommodate the higher momentum results. QCD is assymptotically free, and the perturbative approach
leading to Eq. (\ref{tff}) must be right for high enough momentum transfer. To
correct the high momentum region without touching the low momentum region is
not easy, and might require new effects, as the one shown in ref.
\cite{Kochelev:2009nz}.

On the other hand, using $Q_{0}=1%
%TCIMACRO{\unit{GeV}}%
%BeginExpansion
\operatorname{GeV}%
%EndExpansion
$ as matching point, the high momentum results are well reproduced, but with a
too small slope, and the lower momentum data are over estimated. Here we are
in a more confortable situation from the point of view of QCD. In order to
cure the divergences in the integrand of Eq.(\ref{tff}) we have introduced a
cutoff mass $M$ in the definition of the $\pi TFF$ in terms of the $\pi DA$, recall
Eq. (\ref{tff_R}). This procedure incorporates effects of twist three
operators at low energy. The set of twist three distribution amplitudes for
pions are $\left\langle 0\right\vert \bar{d}\left(  0\right)  i\,\gamma
_{5}\,\tau^{i}\,u\left(  z\right)  \left\vert \pi^{j}\right\rangle ,$
$\left\langle 0\right\vert \bar{d}\left(  0\right)  \,\sigma_{\alpha\beta
}\,\gamma_{5}\,\tau^{i}u\left(  z\right)  \left\vert \pi^{j}\right\rangle $
and $\left\langle 0\right\vert \bar{d}\left(  0\right)  \,\sigma_{\alpha\beta
}\,\gamma_{5}\,\tau^{i}\,G_{\mu\nu}\left(  y\right)  \,u\left(  z\right)
\left\vert \pi^{j}\right\rangle .$ Some of the effects of these operators are
included in the modification proposed by Radyushkin, but there can be
additional effects, like the presence of gluons, which are not incorporated in
his procedure. These additional effects can be introduced by adding to the
lowest order calculation a term proportional to $Q^{-2},$%
\begin{equation}
Q^{2}F(Q^{2})=\frac{\sqrt{2}f_{\pi}}{3}\int_{0}^{1}\frac{dx}{x+\frac{M^{2}%
}{Q^{2}}}\phi_{\pi}(x,Q^{2})+\frac{C_{3}}{Q^{2}}. \label{tff_R_T3}%
\end{equation}
The effect of $C_{3}$ decreases rapidly with $Q^{2}$, and therefore we
consider $C_{3}$ constant in $Q^{2}$. Even though this constant is connected
to twist three operators, its QCD evolution is unimportant here.

Following the same ideas developped in Section \ref{pTFF}, we change the
matching condition, Eq (\ref{cont}) in order to introduce the effect of the
new term:%

\begin{equation}
\frac{\sqrt{2}f_{\pi}}{3}ln\frac{Q_{0}^{2}+M^{2}}{M^{2}}+\frac{C_{3}}%
{Q_{0}^{2}}=\frac{F\left(  0\right)  \,Q_{0}^{2}}{1+a\frac{Q_{0}^{2}}%
{m_{\pi^{0}}^{2}}}. \label{cont3}%
\end{equation}
With $Q_{0}=1%
%TCIMACRO{\unit{GeV}}%
%BeginExpansion
\operatorname{GeV}%
%EndExpansion
$, this equation allows to determine $M,$ once we have fixed the value of
$C_{3}.$

We consider three scenarios, which correspond to a contribution from the twist
three term to the form factor at $Q_{0}=1%
%TCIMACRO{\unit{GeV}}%
%BeginExpansion
\operatorname{GeV}%
%EndExpansion
$ of 10\% ($C_{3}=0.99\,10^{-2}%
%TCIMACRO{\unit{GeV}}%
%BeginExpansion
\operatorname{GeV}%
%EndExpansion
^{3}),$ 20\% ($C_{3}=1.98\,10^{-2}%
%TCIMACRO{\unit{GeV}}%
%BeginExpansion
\operatorname{GeV}%
%EndExpansion
^{3})$ and 30\% ($C_{3}=2.98\,\ 10^{-2}%
%TCIMACRO{\unit{GeV}}%
%BeginExpansion
\operatorname{GeV}%
%EndExpansion
^{3})$. If we fix\footnote{A dimensionless definition of the parameter $C_{3}$
could be $C_{3}=a_{3}\,\left(  m_{\pi^{0}}^{2}/a\right)  \,f_{\pi},$ with $a$
the experimental parameter given in Eq. (\ref{tff_EXP}). Therefore,
$C_{3}=1.98\,10^{-2}%
%TCIMACRO{\unit{GeV}}%
%BeginExpansion
\operatorname{GeV}%
%EndExpansion
^{3}$ corresponds to $\ a_{3}=0.27.$ In the Appendix we give the value for
$C_{3}$ obatined from the triangle diagram.} $C_{3}=1.98\cdot10^{-2}%
%TCIMACRO{\unit{GeV}}%
%BeginExpansion
\operatorname{GeV}%
%EndExpansion
^{3},$ which implies that this term is responsible for a 20\% of the value of
the $\pi TFF$ at $Q_{0}=1%
%TCIMACRO{\unit{GeV}}%
%BeginExpansion
\operatorname{GeV}%
%EndExpansion
,$ the experimental error in $a$ produces a variation of $M=0.620\pm0.050%
%TCIMACRO{\unit{GeV}}%
%BeginExpansion
\operatorname{GeV}%
%EndExpansion
$. To check the stability of the $M$ value in relation to $C_{3},$ we fix
$a=0.032$ and vary $C_{3}$ from $C_{3}=0.99\,10^{-2}%
%TCIMACRO{\unit{GeV}}%
%BeginExpansion
\operatorname{GeV}%
%EndExpansion
^{3}$ to $C_{3}=2.98\,\ 10^{-2}%
%TCIMACRO{\unit{GeV}}%
%BeginExpansion
\operatorname{GeV}%
%EndExpansion
^{3}.$ This produces a variation of $M$ given by $M=0.620\pm0.070%
%TCIMACRO{\unit{GeV}}%
%BeginExpansion
\operatorname{GeV}%
%EndExpansion
.$ The indeterminacy in $M$ is in all cases of the order of 10\%, which is of
the same order of the relative error in the experimental determination of $a$.
The value of $M$ is sufficiently stable to confirm the consistency of the
procedure. For the numerical analysis we have chosen $\Lambda_{QCD}=0.226%
%TCIMACRO{\unit{GeV}}%
%BeginExpansion
\operatorname{GeV}%
%EndExpansion
,$ $Q_{0}=1.%
%TCIMACRO{\unit{GeV}}%
%BeginExpansion
\operatorname{GeV}%
%EndExpansion
$, $a=0.032$ and $C_{3}=1.98\,10^{-2}%
%TCIMACRO{\unit{GeV}}%
%BeginExpansion
\operatorname{GeV}%
%EndExpansion
^{3}$ which corresponds to a 20\% contribution of twist 3 at $Q_{0}.$ This set
of parameters implies $M=0.620%
%TCIMACRO{\unit{GeV}}%
%BeginExpansion
\operatorname{GeV}%
%EndExpansion
$.

\begin{figure}[ptb]
\begin{center}
\includegraphics[
height=8.8458cm,
width=10.5987cm
]{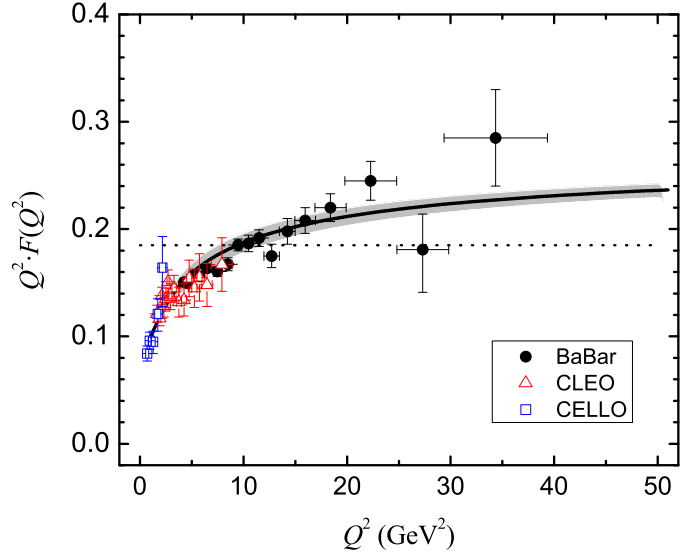}
\end{center}
\caption{Calculation of the transition form factor via the $\pi DA$ with
$M=0.620\operatorname{GeV}$, $a=0.032,\ C_{3}=1.98\,10^{-2}\operatorname{GeV}%
^{3}$ and defining the matching point at $Q_{0}=1\operatorname{GeV}$
(full-line) compared with the experimental data. The gray region gives the
indeterminacy of the results due to the indeterminacy on $C_{3}.$}%
\label{Q01}%
\end{figure}

\begin{figure}[ptb]
\begin{center}
\includegraphics[
height=8.8458cm,
width=10.5987cm
]{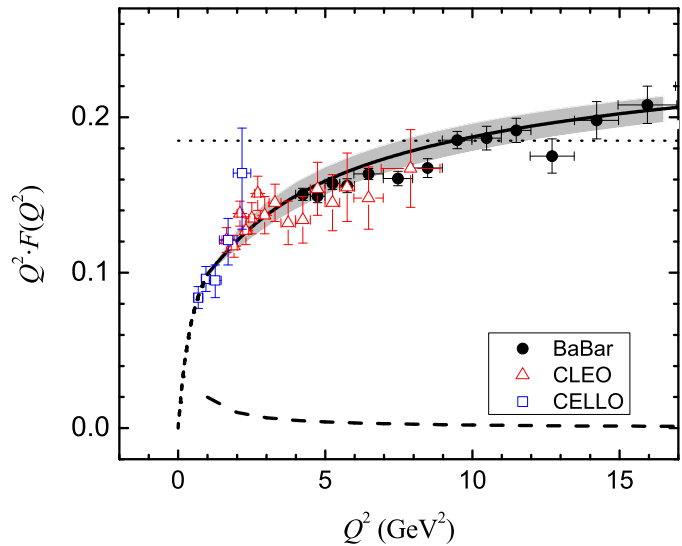}
\end{center}
\caption{Same as in Fig \ref{Q01} for the $Q^{2}$ region under
$17\operatorname{GeV}^{2}.$ The dashed-line corresponds to the $C_{3}$
contribution. The short dashed-line is the experimental parametrization
(\ref{tff_EXP}) for $Q^{2}<1\operatorname{GeV}^{2}.$}%
\label{Q01_0}%
\end{figure}

In Fig. \ref{Q01} and \ref{Q01_0} we show the $\pi TFF$ calculated using the
evolved $\pi DA$ compared with the data of \cite{Behrend:1990sr,
Gronberg:1997fj, :2009mc}. In Fig \ref{Q01_0} we have used the experimental
parametrization (\ref{tff_EXP}) for $Q^{2}<1%
%TCIMACRO{\unit{GeV}}%
%BeginExpansion
\operatorname{GeV}%
%EndExpansion
^{2}$ (short dashed line). For $Q^{2}>1%
%TCIMACRO{\unit{GeV}}%
%BeginExpansion
\operatorname{GeV}%
%EndExpansion
^{2}$, the full line corresponds our calculation using Eq. (\ref{tff_R_T3})
together with the evolved $\pi DA.$ The gray region corresponds to the effect
due to the different contributions of the $C_{3}$ term: the upper limit
corresponds to $C_{3}=0.99\,10^{-2}%
%TCIMACRO{\unit{GeV}}%
%BeginExpansion
\operatorname{GeV}%
%EndExpansion
^{3}$ (10\% of twist 3 contribution at $Q^{2}=1%
%TCIMACRO{\unit{GeV}}%
%BeginExpansion
\operatorname{GeV}%
%EndExpansion
);$ the lower limit corresponds to $C_{3}=2.98\,10^{-2}%
%TCIMACRO{\unit{GeV}}%
%BeginExpansion
\operatorname{GeV}%
%EndExpansion
^{3}$ (30\% of twist 3 contribution at $Q^{2}=1%
%TCIMACRO{\unit{GeV}}%
%BeginExpansion
\operatorname{GeV}%
%EndExpansion
)$; the dashed curve in Fig. \ref{Q01_0} corresponds to the contribution
coming from the $C_{3}$ term for our central value, $C=1.98\cdot10^{-2}%
%TCIMACRO{\unit{GeV}}%
%BeginExpansion
\operatorname{GeV}%
%EndExpansion
^{3}.$ We observe that its contribution is absolutely negligible for $Q^{2}>5%
%TCIMACRO{\unit{GeV}}%
%BeginExpansion
\operatorname{GeV}%
%EndExpansion
^{2}.$ Nevertheless, its influence in the matching point is important for a
good description of the low $Q^{2}$ data. Looking at the whole of the Figs.
\ref{Q01} and \ref{Q01_0}, we observe that we have a good determination of the
experimental data in the whole experimental region. 

\section{Conclusions}

In the previous sections we have developed a formalism to describe the $\pi
TFF$ on all experimentally accessible range, and hopefully beyond. The
formalism is based on a two energy scale description. The formulation in the
low energy scale is non perturbative, while that of the high energy scale is
fundamentally a QCD based perturbative formulation. The two descriptions are
matched at an energy scale $Q_{0}$ hereafter called hadronic scale. This
scheme has been applied to described the parton and generalized parton
distributions with notable
success\cite{Traini:1997jz,Scopetta:1997wk,Scopetta:1999ud,Theussl:2002xp}. In
order to understand the finesse of the BaBaR data we have forced us to
introduce additional ingredients to this scheme which we next summarize.

We have taken for the low energy description of the $\pi TFF$ a parameterized
description of the data. This has been done to avoid model dependence in a
region ($Q\sim1$ GeV), where models might start to break down. The high energy
description incorporates the following important physical ingredients:
\textit{i)} a mass cut-off in the definition of the $\pi TFF$ from the $\pi
DA,$ $M,$ \cite{Radyushkin:2009zg} which, interpreted from the point of view
of constituent models, takes into account the constituent mass, transverse
momentum effects and also higher twist effects; \textit{ii)} a flat $\pi DA$
at low energies in line with chiral models
\cite{Praszalowicz:2001wy,RuizArriola:2002bp,Courtoy:2007vy,RuizArriola:2003bs}
and recent proposals \cite{Radyushkin:2009zg,Polyakov:2009je} ; \textit{iii)}
eventually an additional twist three term into the definition of the $\pi TFF$
in the high energy description parameterized by a unique constant $C_{3}$;
\textit{iv)} the two descriptions have to match at an energy scale $Q_{0}$;
this scale is universal and should be the same for all observables. Though our scheme
is not based in the use a hadron model, we have presented in the
Appendix a simple model calculation to illustrate  the contents of the procedure.

We have seen that, as a consequence of soft pion theorems, the $\pi DA$ and
$\pi VPD$ are equal at low energies up to the matching scale $Q_{0}$.
Moreover, chiral models indicate that in the chiral limit both distributions
are flat, more precisely, $\phi_{\pi}\left(  x,Q_{0}\right)  =q_{V}\left(
x,Q_{0}\right)  =1.$ The flatness of the $\pi VPD,$ which\ implies that the
probability of finding a quark of any relative momentum fraction $x$ is the
same, is justified by the fact that in the chiral limit quarks and pions are
massless and therefore there is no mass scale to which the momentum
distribution can attach. The different tipe of evolution followed by these two
distributions hides the relation between them. Nevertheles, the equality
between the $\pi DA$ and the $\pi VPD$\ at $Q_{0}$ scale allows us to
determine the matching scale using the parton distributions. From previous
studies \cite{Davidson:1994uv, Broniowski:2007si, Courtoy:2008nf} we can fix
this scale at $Q_{0}=0.355%
%TCIMACRO{\unit{GeV}}%
%BeginExpansion
\operatorname{GeV}%
%EndExpansion
.$ Nevertheless, more elaborated studies
\cite{Scopetta:1997wk,Scopetta:1999ud,Noguera:2005cc} indicates that this
matching scale cannot be one in which the constituent quarks are purely
valence but require additional sea and gluon components. Theses studies push
the matching scale at $Q_{0}\sim1$GeV.

Using $Q_{0}=0.355%
%TCIMACRO{\unit{GeV}}%
%BeginExpansion
\operatorname{GeV}%
%EndExpansion
$ as matching point, the experimental results are well reproduce up to $15%
%TCIMACRO{\unit{GeV}}%
%BeginExpansion
\operatorname{GeV}%
%EndExpansion
^{2},$ but it is difficult to accommodate the higher momentum results. If we
adopt this philosophy, the BaBar data, if confirmed, might imply some new type of
contributions \cite{Kochelev:2009nz}. Using
$Q_{0}=1%
%TCIMACRO{\unit{GeV}}%
%BeginExpansion
\operatorname{GeV}%
%EndExpansion
$ as matching point, the higher momentum results are well reproduced, but with
a slope which is too small, and the lower momentum data are overestimated.
Nevertheless this last situation can be corrected in a natural way if we
assume some additional contribution from twist three.

The calculations shown prove that the BaBar
results can be accommodated in our scheme, which only uses standard QCD
ingredients and low energy data. It must be emphasized that in order to have a
good description higher twist effects are important as the modification from
Eq. (\ref{tff}) to Eq. (\ref{tff_R_T3}) signals. It must be also noted that
the matching scale is as high as $1$ GeV, a feature which was also the case in
the description of parton distributions when precision was to be attained.
With these ingredients our calculation shows an excellent agreement with the
data and points out that maybe the average value of the highest energy data
point is too large, a conclusion reached by other analyses
\cite{Mikhailov:2009sa,Dorokhov:2009zx}.

The relative high value of $M=0.620%
%TCIMACRO{\unit{GeV}}%
%BeginExpansion
\operatorname{GeV}%
%EndExpansion
$ can be understood on the basis that it includes the constituent quark mass,
the mean value of the transverse quark momentum and other higher twist
contributions. The $C_{3}$ term is relatively small. Its effect is to reduce
the value of the contribution to the $\pi TFF$ of twist two only for low
$Q^{2}<5$ GeV$^{2}$. Our results are very stable with respect to variations of
these parameters.

Let us conclude by stressing that we have developed a formalism to describe
the $\pi TFF$ based on the philosophy that two descriptions, a perturbative
and a non perturbative can be matched in a physically acceptable manner at a
certain scale $Q_{0}$. The idea of the formalism is that one can use models or
effective theories to describe the non perturbative sector while QCD to
describe the perturbative one. In here we have preferred to use data for the
low energy sector to avoid model dependence, but in other observables, and
specially in predictions, models can be used for estimates. Moreover, in order
to describe the data we have seen that a matching scale of $1$ GeV was favored
over the one of $0.355$ GeV that we would obtain if only valence constituents
were assumed to describe low energies. This implies that the description of
high precision data requires sophisticated effective theories (models) at low
energies. We have also discovered that higher twist effects (parametrized in
our case by $M$ and $C_{3}$) are small but crucial in order to attain good
descriptions. Finally we can assert that our description of the data is
excellent all over although points out to a too large average value of the
highest $Q$ points in the data.

\section*{Acknowledgements}

One of us VV would like to thank Nikolai Kochelev for useful discussions. We
thank the authors of JaxoDraw for making drawing diagrams an easy task
\cite{Binosi:2003yf}. This work was supported in part by HadronPhysics2, a
FP7-Integrating Activities and Infrastructure Program of the European
Commission under Grant 227431, by the MICINN (Spain) grant FPA2007-65748-C02-01
and by GVPrometeo2009/129.

\appendix{}

\section{The triangle diagram.}

 The easiest model to evaluate the $\pi TFF$ is to use the triangle diagram
defined in Fig. \ref{Trianglediagrams}. There are two diagrams, the one
depicted an another with the photons exchanged. The amplitude for the process
can be defined as%
\begin{equation}
\mathcal{T}=4\,\pi\,\alpha\,\varepsilon^{\mu}\,\varepsilon^{\prime\nu}%
\,q^{\pi}\,q^{\prime\sigma}\,\varepsilon_{\mu\nu\rho\sigma}\,F\left(
q,q^{\prime}\right)
\end{equation}
In a straightforward calculation\footnote{A detailed calculation of this process is shown in ref. \cite{Itzykson:1980rh}.
We need  to replace there  $e^{2}$ by $N_{c}\left(  e_{u}%
^{2}-e_{d}^{2}\right)  =e^{2},$ since in our case we have quarks in the loop .},
\begin{equation}
F(q^{2},q^{\prime2})=8\,m_{q}\,g_{\pi qq}\,I(m,q^{2},q^{\prime2}),
\label{tff_NJL}%
\end{equation}
where
\begin{equation}
I(m,q^{2},q^{\prime2})=i\int\frac{d^{4}\ell}{\left(  2\pi\right)  ^{4}}%
\frac{1}{\left(  \ell^{2}-m_{q}^{2}+i\epsilon\right)  \left[  \left(
q-\ell\right)  ^{2}-m_{q}^{2}+i\epsilon\right]  \left[  \left(  q^{\prime
}+\ell\right)  ^{2}-m_{q}^{2}+i\epsilon\right]  }\ ,
\end{equation}
$q_{\mu}$ and $q_{\mu}^{\prime}$ are the photon momenta, which, in this
expression, can be real or virtual photons. Here, $m_{q}$ is the constituent
quark mass, $g_{\pi qq}$ the pion-quark coupling constant, $q$ and $q^{\prime
}$ the photons momenta. The small virtuality of one of the photons implies
that $q^{\prime2}=0,$ and we define $F(Q^{2})=F
(q^{2},0)$ with $Q^{2}=-q^{2}.$ In the chiral limit, $m_{\pi}=0,$ we can use
the Goldberger-Treiman relation, $g_{\pi qq}/m_{q}=\sqrt{2}/f_{\pi},$
obtaining%
\begin{equation}
F(Q^{2})=\frac{\sqrt{2}\,m_{q}^{2}}{2\,\pi^{2}\,f_{\pi}}\int
_{0}^{1}dx\frac{1}{Q^{2}x\sqrt{1+2\,u}}\log\left\vert \frac{1+u+\sqrt{1+2\,u}%
}{1+u-\sqrt{1+2\,u}}\right\vert \label{tff_NJL_1}%
\end{equation}
where $u=2m_{q}^{2}/(Q^{2}x^{2}).$

The behavior of the form factor for small $Q^{2}$\ values follows from the
previous equation,%

\begin{equation}
F(Q^{2})\underset{Q^{2}\rightarrow0}{\longrightarrow}\frac{\sqrt
{2}}{4\pi^{2}f_{\pi}}\left(  1-\frac{Q^{2}}{12\,m_{q}^{2}}\right)
\end{equation}
Therefore, we have $F(0)=\sqrt{2}/\left(  4\pi^{2}f_{\pi}\right)  =0.273%
%TCIMACRO{\unit{GeV}}%
%BeginExpansion
\operatorname{GeV}%
%EndExpansion
^{-1},$ and ,from Eq. (\ref{ExpFit}), $m_{q}=m_{\pi}/\sqrt{12\,a}\simeq0.220%
%TCIMACRO{\unit{GeV}}%
%BeginExpansion
\operatorname{GeV}%
%EndExpansion
.$

We can also look for the asymptotic behavior. The denominator of Eq.
(\ref{tff_NJL_1}) can be developped in the form $x\sqrt{1+2\,u}\sim
x+2m_{q}^{2}/(Q^{2}x).$ Comparing this expression with Eq. (\ref{tff_R}), we
have $M=\sqrt{2m_{q}^{2}/x}\sim2m_{q}=0.44%
%TCIMACRO{\unit{GeV}}%
%BeginExpansion
\operatorname{GeV}%
%EndExpansion
.$ Performing the integral in Eq. (\ref{tff_NJL_1}) we have
\begin{equation}
Q^{2}F(Q^{2})\ \underset{Q^{2}\rightarrow\infty}{\longrightarrow
}\ \frac{m_{q}^{2}}{2\,\sqrt{2}\,\pi^{2}\,f_{\pi}}\log\frac{Q^{2}}{m_{q}^{2}%
}\left[  \log\frac{Q^{2}}{m_{q}^{2}}+4\,\frac{m_{q}^{2}}{Q^{2}}+...\right]
...\,\ .
\end{equation}
Therefore, the model calculation gives a wrong asymptotic behavior. We have
$Q^{2}F(Q^{2})\sim0.54%
%TCIMACRO{\unit{GeV}}%
%BeginExpansion
\operatorname{GeV}%
%EndExpansion
$ for $Q^{2}\sim30%
%TCIMACRO{\unit{GeV}}%
%BeginExpansion
\operatorname{GeV}%
%EndExpansion
^{2}$ which is a very large result, and the form factor grows as $\left(  \log
Q^{2}/m_{q}^{2}\right)  ^{2}$ for large $Q^{2}$ values. Comparing this
expression with Eq. (\ref{tff_R_T3}), we obtain $C_{3}=\sqrt{2}m_{q}^{4}%
\log\left(  Q_{0}^{2}/m_{q}^{2}\right)  /\pi^{2}f_{\pi}\sim0.78\,10^{-2}%
%TCIMACRO{\unit{GeV}}%
%BeginExpansion
\operatorname{GeV}%
%EndExpansion
^{3},$ which is consistent with the values for $C_{3}$ used in the text.

\end{document}